%% file: ms.tex
\documentclass[sigplan,screen,10pt]{acmart}
\startPage{1}

\setcopyright{rightsretained}
\acmPrice{}
\acmDOI{10.1145/3359619.3359744}
\acmYear{2019}
\copyrightyear{2019}
\acmISBN{}
\acmConference[Dynamic Languages Symposium]{Proceedings of the 15th ACM SIGPLAN International Symposium on Dynamic Languages}{2019}{Preprint}
\acmBooktitle{Proceedings of the 15th ACM SIGPLAN International Symposium on Dynamic Languages (DLS '19), October 20, 2019, Athens, Greece}



\usepackage{booktabs}  
\usepackage{listings}
\usepackage{xcolor}
\usepackage{mathpartir}
\usepackage{amsmath,xspace}
\usepackage{ottlayout}
\usepackage{tikz}
\usetikzlibrary{shapes,positioning,arrows,calc}

\newcommand{\MkEnv}{\ensuremath{\pirI{MkEnv}}\xspace}
\newcommand{\MkArg}{\ensuremath{\pirI{MkArg}}\xspace}
\newcommand{\Call}{\ensuremath{\pirI{Call}}\xspace}
\newcommand{\MkClosure}{\ensuremath{\pirI{MkClosure}}\xspace}
\newcommand{\StVar}{\ensuremath{\pirI{StVar}}\xspace}
\newcommand{\LdVar}{\ensuremath{\pirI{LdVar}}\xspace}
\newcommand{\LdArg}{\ensuremath{\pirI{LdArg}}\xspace}
\newcommand{\Force}{\ensuremath{\pirI{Force}}\xspace}
\renewcommand{\Phi}{\ensuremath{\pirI{Phi}}\xspace}
\newcommand{\Deopt}{\ensuremath{\pirI{Deopt}}\xspace}
\newcommand{\LdFun}{\ensuremath{\pirI{LdFun}}\xspace}
\newcommand{\LdConst}{\ensuremath{\pirI{LdConst}}\xspace}
\newcommand{\Branch}{\ensuremath{\pirI{Branch}}\xspace}
\newcommand{\get}[3]{\ensuremath{#1_{#2,#3}}\xspace}

\begin{document}

\title{R Melts Brains}
\subtitle{An IR for First-Class Environments and Lazy Effectful Arguments}

\author{Olivier Fl\"uckiger}
\affiliation{\institution{Northeastern University}}
\author{Guido Chari}
\affiliation{\institution{Czech Technical University}}
\author{Jan Je\v{c}men}
\affiliation{\institution{Czech Technical University}}
\author{Ming-Ho Yee}
\affiliation{\institution{Northeastern University}}
\author{Jakob Hain}
\affiliation{\institution{Northeastern University}}
\author{Jan Vitek}
\affiliation{\institution{Northeastern / Czech Technical U.}}

\renewcommand{\shortauthors}{Fl\"uckiger, Chari, Je\v{c}men, Yee, Hain, Vitek}
\newcommand{\R}{{\bf\textsf{\v{R}}}\xspace}

\begin{abstract}
The R programming language combines a number of features considered hard to
analyze and implement efficiently: dynamic typing, reflection, lazy
evaluation, vectorized primitive types, first-class closures, and extensive
use of native code.  Additionally, variable scopes are reified at runtime as
first-class environments.  The combination of these features renders most
static program analysis techniques impractical, and thus, compiler
optimizations based on them ineffective.  We present our work on \PIR, an
intermediate representation with explicit support for first-class
environments and effectful lazy evaluation.  We describe two dataflow
analyses on \PIR: the first enables reasoning about variables and their
environments, and the second infers where arguments are evaluated.
Leveraging their results, we show how to elide environment creation and
inline functions.
\end{abstract}

\lstset{language=R}
\definecolor{LightGray}{rgb}{.92,.92,.92}
\definecolor{Gray}{rgb}{.3,.3,.3}
\definecolor{DarkGray}{rgb}{.5,.5,.5}
\lstset{ %
  columns=flexible,
  captionpos=b,
  frame=single,
  framerule=0pt,
  tabsize=2,
  belowskip=0.5em,
  backgroundcolor=\color{LightGray},
  basicstyle=\small\ttfamily,
  emphstyle=,
  keywordstyle=,
  commentstyle=\color{Gray}\em,
  stringstyle=\color{Gray}
}
\lstdefinestyle{R}{ %
  language=R,
  morekeywords={assign, delayedAssign},
  deletekeywords={env, equal, c, runif, trace, args, exp, t, all},
  breaklines=true
}
\lstdefinestyle{Rin}{ %
  style=R,
  breaklines=false
}

\newcommand{\eg}{\emph{e.g.},\xspace}
\newcommand{\ie}{\emph{i.e.},\xspace}

\newcommand{\PIR}{\textsf{PIR}\xspace}
\newcommand\pirI[1]{\mathtt{#1}}

\input{pir_}

\renewcommand{\ottpremise}[1]{\premiseSTY{#1}}
\renewcommand{\ottusedrule}[1]{\usedruleSTY{#1}}
\renewcommand{\ottdrule}[4][]{\druleSTY[#1]{#2}{#3}{#4}}
\renewenvironment{ottdefnblock}[3][]{\defnblockSTY[#1]{#2}{#3}}{\enddefnblockSTY}
\renewcommand{\c}[1]{{\lstinline[style=Rin]!#1!}\xspace}

\maketitle

\section{Introduction}
\label{sec:intro}

The R language~\citep{rco19} presents interesting challenges for
implementers.  R is a dynamic imperative language with vectorized
operations, copy-on-write of shared data, a call-by-need evaluation
strategy, context-sensitive lookup rules, multiple dispatch, and first-class
closures.  A rich reflective interface and a permissive native interface
allow programs to inspect and modify most of R's runtime structures.  This
paper focuses on the interplay of first-class, mutable environments and lazy
evaluation. 
In particular, we focus on their impact on compiler optimizations.

One might see the presence of \c{eval} as the biggest obstacle for static
reasoning.  With \c{eval}, text can be turned to code and perform arbitrary
effects.  However, the expressive power of \c{eval} can be constrained by
careful language design. Julia, for instance, has a reflective interface
that does not hamper efficient compilation~\cite{oopsla18a}.  Even an
unconstrained \c{eval} is bound by what the language allows; for example,
most programming languages do not allow code to delete a variable.  Not so
in R.  Consider one of the most straightforward expressions in any language,
variable lookup:

\begin{lstlisting}
  f <- function(x) x
\end{lstlisting}

\noindent In most languages, it is compiled to a memory or register access.
From the point of view of a static analyzer, this expression usually leaves
the program state intact.  Not so in R.
Consider a function doubling its argument:
\begin{lstlisting}
  g <- function(x) x+x
\end{lstlisting}

\noindent In most languages, a compiler can assume it is equivalent to
\c{2*x} and generate whichever code is most efficient. At the very least,
one could expect that both lookups of \c{x} resolve to the same variable.
Not so in R.

Difficulties come from two directions at once.  R variables are bound in
environments, which are first-class values that can be modified.  In
addition, arguments are evaluated lazily; whenever an argument is accessed
for the first time, it may trigger a side-effecting computation -- which could
modify any environment.  Consequently, to optimize the body of a function, a
compiler must reason about effects of the functions that it calls,
as well as the effects from evaluating its arguments.
In the above example, \c{`+`} could walk up the call stack and
delete the binding for variable~\c x.  One could also call \c g with an
expression that deletes \c x and causes the second lookup of \c x to fail.
While unlikely, a compiler must be ready for it. Considering these
examples in combination with \c{eval}, it is impossible to statically resolve the binding
structure of R programs.  Unsurprisingly, existing
implementations resort to dynamic techniques to optimize code~\citep{tal12,
  wan14, vee14, sta16}.

The contribution of this paper is the design of \PIR, an intermediate
representation (IR) for R programs with explicit support for environments
and lazy evaluation. 
\PIR is a static single assignment (SSA)~\citep{ros88} code
format inspired by our experience with the bytecode of the GNU R
reference implementation, earlier work on FastR~\citep{vee14}, the sourir IR
we developed to model speculative optimizations~\citep{popl18}, and an
earlier attempt to optimize R using LLVM. 
In our experience, some of the
most impactful optimizations are high-level ones that require understanding
how values are used across function boundaries. We found that the GNU R
bytecode~\citep{tie19} was too high level; it left too many of the
operations implicit.  In contrast, we found LLVM's IR~\citep{lat04} too low
level for easily expressing some of our target optimizations.

\PIR is part of \R, a new just-in-time compiler for the R language.  To
motivate its need, we start with background on R and on related efforts in
\autoref{sec:r-primer}. We give an informal overview of \PIR in
\autoref{sec:approach}. Then, \autoref{sec:pir} details \PIR and presents
two transformation passes.  The first, scope resolution, statically resolves
bindings, and the second, promise inlining, removes lazy argument
evaluation.  Finally, \autoref{sec:results} illustrates how \PIR helps
\R\footnote{Pronounced like a trilled ``r'', the sound one makes upon
  realizing that arguments can modify the environment of the function they
  are given to.}  reduce overheads. Our compiler is not complete and we are
not yet able to run at competitive speed, so the results should be
considered preliminary.
\R is available at
\url{https://github.com/reactorlabs/rir}.

\section{Background}\label{sec:r-primer}

This section describes key properties of environments and promises, and
discusses work that deals with similar issues.

\subsection{Environments in R}

Inspired by Scheme and departing from its predecessor S, R adopted a lexical
scoping discipline~\cite{gen00}. Variables are looked up in a list of
environments. Consider this snippet:

\begin{lstlisting}
  g <- function(a) { 
      f <- function() x+y         
      if (a) x <- 2
      f() 
  }
  y <- 1
\end{lstlisting}

\noindent The evaluation of \c{x+y} requires finding \c x in the enclosing
environment of the closure \c{f}, and \c y at the top level.  It is worth
pointing out that, while R is lexically scoped, the scope of a free variable
cannot be resolved statically. For instance, \c x will only be in scope in
\c g if the argument \c a evaluates to true.  

R uses a single namespace for functions and variables. Environments are used
to hold symbols like \c +.  While primarily used for variables, environments
can also be created explicitly, \eg to be used as hashmaps.  Libraries are
loaded by the \c{attach()} function that adds an environment to the list of
environments.  A number of operations allow interaction with environments:
\c{environment()} accesses the current environment; \c{ls(...)} lists bound
variables; \c{assign(...)} adds or modifies a binding; and \c{rm(...)}
removes variables from an environment.  R has functions to walk the
environment chain: \c{parent.frame()} returns the environment associated
with the caller's call frame and \c{sys.frame(...)}  provides access to the
environment of any frame on the call stack. In R, frames represent function
invocations and they have references to environments. Consider this code:

\begin{lstlisting}
  f <- function() get("x", envir=parent.frame())
  g <- function() {x <- "secret"; f()}
\end{lstlisting}

\noindent Function \c{f} uses reflection to indirectly access
\c g's environment.  This illustrates that any callee may access (and
change) the caller environment.

\subsection{Laziness in R}

Since its inception, R has adopted a call-by-need evaluation strategy (also
called lazy evaluation).  Each expression passed as argument to a function
is wrapped in a \emph{promise}, a thunk that packages the expression, its
environment, and a slot to memoize the result of evaluating the expression.  A promise is only
evaluated when its value is needed.  Consider a
function that branches on its second argument:

\begin{lstlisting}
  f <- function(a, b) if(b) a
\end{lstlisting}

\noindent A call \c{f(x<-TRUE,x)} creates two promises, one for the
assignment \c{x<-TRUE}, and one to read \c x.  One could expect this call to
return \c{TRUE}, but this is not so.  The condition is evaluated before
variable \c x is defined, causing an error to be reported.  Combined with
promises, the \c{sys.frame} function allows non-local access to environments
during promise evaluation:

\begin{lstlisting}
  f <- function() sys.frame(-1)
  g <- function(x) x
  g(f())
\end{lstlisting}

\noindent Here \c{g} receives promise \c{f()} as argument.  When the promise
is forced, there will be three frames on the stack: frame~0 is the global
scope, frame~1 is \c{g}'s, and frame~2 is \c{f}'s frame.

\begin{figure}[H]\small\begin{tikzpicture}[stack/.style={
   rectangle split, rectangle split parts=3, draw, anchor=west, text width=5cm}]

\node [stack] (t) {
  \nodepart{one}
  \c{0:  g(f())}
  \nodepart{two}
  \c{1:  x}
  \nodepart{three}
  \c{2:  sys.frame(-1)}
};

\end{tikzpicture}
\end{figure}

\noindent
During promise evaluation, \c{parent.frame} refers to the frame where the
promise was created (frame~0 in this example, as promise \c{f()} occurs at
the top level).  But, \c{sys.frame(-1)} accesses a frame by index,
ignoring lexical nesting, thus extracting the environment of the forcing
context, \ie the local environment of \c g at frame~1.

We leave the reader with a rather amusing brain twister.  
R has context-sensitive lookup rules for variables in call position.
Variables that are not bound to functions are skipped:

\begin{lstlisting} 
  f <- function(c) {c(1, 2) + c}
  f(3)
\end{lstlisting}
 
\noindent
The lookup of \c c in \c{c(1,2)} skips the argument \c c, since it is not a
function.  Instead, primitive \c{c()} is called to construct a vector.  The
second read of \c c is not in call position, thus it returns argument \c c,
\c{3} in this case.  The result is the vector \c{[4,5]} as addition is
vectorized.  Now, consider the following variation:

\begin{lstlisting}
  bad <- function() rm(list="c", envir=sys.frame(-1))
  f(bad())
\end{lstlisting}

\noindent
This time evaluation ends with an error as we try to add a vector and a
function.  Evaluation of \c{c(1,2)} succeeds and returns a vector.  But,
during the lookup of \c c for that call, R first encounters the argument \c
c. In order to check if \c c is bound to a closure, it evaluates the
promise, causing \c{bad()} to delete the argument from the environment. On
the second use of \c c, the argument has been removed and a function object,
\c c, is returned.

\subsection{Related Work}

R has one reference implementation, GNU R, and several alternative 
implementations.  
GNU R includes a bytecode compiler with a small number of carefully tuned
optimizations~\cite{tie19}.  
Unlike ours, GNU R's bytecode implicitly
assumes the presence of an environment for every function application.  
Variable lookup, in the
worst case, requires inspecting all bindings of each environment in scope.
To mitigate the lookup cost, GNU R caches bindings when safe.  
FastR's first version featured a type-specializing tree interpreter that
outperformed GNU R~\cite{vee14}. 
It split environments into a statically known part (represented by arrays with
constant-time accesses) and extensions that could grow and shrink at runtime.
Environments were marked dirty whenever a reflective operation modified them.  
The second version of FastR uses Truffle for specialization and Graal for code
generation~\cite{wur13,sta16}. 
Graal's intermediate representation is general purpose~\cite{dub13}.  
FastR speculatively specializes the code based on profile-driven
global assumptions. 
For instance, functions exhibiting a runtime stable binding structure are
compiled under that assumption. 
The compiler elides environments and stores variables on the stack. 
Code is added to detect violation of assumptions and trigger deoptimization.  
Type specialization was also used in the ORBIT project, an attempt at extending
GNU R with a type specializing bytecode interpreter~\cite{wan14}. On the other
hand, the Riposte compiler tried to speed up R by recording execution traces for
vector operations~\citep{tal12}. Riposte performed liveness analysis on the
recorded traces to avoid unnecessary vector creations and parallelize code. None
of these alternatives provides any special treatment for environment bindings.
Our work departs from all these efforts in that we provide explicit support for
environments and promises in the compiler IR.  
This allows us to combine static reasoning (when feasible) with speculative
optimizations (when needed).

Other languages have some of the same features R has but, usually, are more
amenable to compilation.  
Julia resembles R in that it is dynamically typed, reflective, and targets
scientific computing. But, as shown by~\citet{oopsla18a}, it exhibits much
better performance. 
This is due to a combination of careful language design and an implementation
strategy that focuses on type specialization, inlining, and unboxing.  Julia
does not have lazy evaluation, it restricts \c{eval} to execute at the top
level, and limits reflection. 
Another example is JavaScript. 
While it is also dynamic, the only way to add variables to a scope is using \c{eval}, which 
can only do so locally.
\citet{ser18} performs static reasoning on JavaScript by relying on
type specialization and occurrence typing~\cite{Tob10}, as well as rapid atomic
type analysis~\cite{Log10}.  
Whenever types cannot be statically determined, the compiler assumes the most
likely structures ahead of time and relies on speculative guards for soundness.
Smalltalk also features first-class contexts, although adding bindings at
runtime is not supported. 
The Cog VM~\cite{miranda2011cog} maps context objects to the native stack and
materializes contexts on demand when they are reflectively accessed.

\section{An Intermediate Representation for R}\label{sec:approach}

We provide an example-driven explanation of \PIR before the formal introduction.
For readers who prefer a bottom-up explanation, we suggest
starting with \autoref{sec:pir}. We distinguish between source-level R
variables, which we call \emph{variables}, and \PIR local variables, called \emph{registers}. Variables are stored in environments
while the implementation of registers is left
up to the compiler, and reflective access is not provided.

\subsection{Scope Resolution to Lower Variables} 

We start with an example to illustrate how R variables are modeled, and if possible lowered to registers.
We use the following simple function definition:

\begin{lstlisting}
  function() { answer <- 42; answer }
\end{lstlisting}

\noindent The function defines a local variable and returns its value.  It translates to the following \PIR instructions:

 {\ttfamily\small
                 \vskip -0.5em
                  \[
                      \begin{array}{rlll}
                           &   \mathtt{e{ 0 } }   & = &   \pirI{MkEnv}~  ( \, ~  :  \mathsf{G}  )   \\  \,  &   \mathtt{\% 1 }   & = &   \pirI{LdConst}~        [  1  ]~ 42    \\  \,  &        &   &   \pirI{StVar}~     (  {\sf answer }  ,~  \mathtt{\% 1 }  ,~  \mathtt{e{ 0 } }  )   \\  \,  &   \mathtt{\% 3 }   & = &   \pirI{LdVar}~        (  {\sf answer }  ,~  \mathtt{e{ 0 } }  )   \\  \,  &   \mathtt{\% 4 }   & = &   \pirI{Force}~        (  \mathtt{\% 3 }  )~  \mathtt{e{ 0 } }    \\  \,  &        &   &   \pirI{Return}~       (  \mathtt{\% 4 }  )   \\  
                      \end{array}
                  \]
                 \vskip -0.01em
              } 

\noindent First, \MkEnv creates an empty environment nested in $ \mathsf{G} $,
the global environment.  As all values are vectorized, \c{42} is loaded as a
vector of length 1.  \StVar updates environment $ \mathtt{e{ 0 } } $ with a binding for
variable \c{answer}.  Then, \LdVar loads variable \c{answer} again.  As the examples
in \autoref{sec:r-primer} have conveyed, the compiler cannot assume much
about the loaded value.  Because returns are strict in R, the compiler
inserts a \Force instruction to evaluate promises. 
It refers to environment $ \mathtt{e{ 0 } } $ because a promise could reflectively access it.  
We record this fact as a data dependency.  
In general, we use a notation where actual arguments are inside parentheses and
data dependencies outside.  
When \Force is passed a value, rather than a promise, it does nothing.

After translation, the compiler runs a scope resolution pass to lower
variables to registers.  This requires combining an analysis and a
transformation step.  The analysis computes the reaching stores at each
program point. Its results are then used to remove loads.  In the previous
example, the analysis proves that the value referenced by variable \c{answer} in
instruction $ \mathtt{\% 3 } $ originates from $ \pirI{StVar}~     (  {\sf answer }  ,~  \mathtt{\% 1 }  ,~  \mathtt{e{ 0 } }  ) $.  Thus, $ \mathtt{\% 3 } $
can be substituted with $ \mathtt{\% 1 } $.  In case of multiple dominating stores, we
insert a \Phi instruction to combine them into a single register.
Once this load is resolved, the environment is not used anymore, except for a
dead store.  Standard compiler optimizations, such as escape analysis of the
environment and dead store elimination, can now transform this function into:

 {\ttfamily\small
                 \vskip -0.5em
                  \[
                      \begin{array}{rlll}
                           &   \mathtt{\% 1 }   & = &   \pirI{LdConst}~        [  1  ]~ 42    \\  \,  &        &   &   \pirI{Return}~       (  \mathtt{\% 1 }  )   \\  
                      \end{array}
                  \]
                 \vskip -0.01em
              } 

\noindent This version has no loads, stores, or environment and does not
require speculation.

\subsection{Promise Elision}\label{sec:initial-promise-example}
Promises consume heap memory and hinder analysis, since they might have
side effects.  
Therefore, we statically elide them when possible with the following
three steps: first, inline the callee; next, identify where the promise is
evaluated; and last, inline the body of the promise at that location.  To
preserve observable behavior, inlining must ensure that side effects happen
in the correct order.  Consider the following code snippet:

\begin{lstlisting}
  f <- function(b) b
  f(x)
\end{lstlisting}

This snippet translates to the following \PIR instructions.
First we show the creation of closure \c f and its invocation with one promise argument \c x:
\vskip -1em
 {\ttfamily\small
                 \vskip -0.5em
                  \[
                      \begin{array}{rlll}
                          \multicolumn{4}{l}{
                               } \\   &   \mathtt{\% 1 }   & = &   \pirI{MkClosure}~(  {\sf f }  ,~ \mathsf{G} )   \\  \,  &   \mathtt{\% 2 }   & = &   \pirI{MkArg}~    (  {\sf pr0 }  ,~ \mathsf{G} )   \\  \,  &   \mathtt{\% 3 }   & = &   \pirI{Call}~    \mathtt{\% 1 }  ~(  \mathtt{\% 2 }  )~ \mathsf{G}   \\   \multicolumn{4}{l}{  \,  \mathsf{  {\sf pr0 }  } } \\   &   \mathtt{\% 4 }   & = &   \pirI{LdVar}~        (  {\sf x }  ,~ \mathsf{G} )   \\  \,  &   \mathtt{\% 5 }   & = &   \pirI{Force}~        (  \mathtt{\% 4 }  )~ \mathsf{G}   \\  \,  &        &   &   \pirI{Return}~       (  \mathtt{\% 5 }  )   \\   \multicolumn{4}{l}{  
                          }
                      \end{array}
                  \]
                 \vskip -0.01em
              } 
\vskip -0.5em
\noindent The closure is explicitly created by \MkClosure.  Similarly, the
promise $ \mathtt{\% 2 } $ is created by \MkArg from \c{pr0}.  Analogous to \Force, \Call has a data
dependency on the environment because the callee can potentially access it.
The translation of \c{pr0} does not optimize the read of \c x as this would
require the equivalent to an interprocedural analysis.

\noindent Then, the function \c f translates to the following \PIR:
 {\ttfamily\small
                 \vskip -0.5em
                  \[
                      \begin{array}{rlll}
                          \multicolumn{4}{l}{
                               \mathsf{  {\sf f }  } } \\   &   \mathtt{\% 6 }   & = &   \pirI{LdArg}~        ( 0 )   \\  \,  &   \mathtt{e{ 7 } }   & = &   \pirI{MkEnv}~  (  {\sf b }   \ottsym{=}   \mathtt{\% 6 }  ~  :  \mathsf{G}  )   \\  \,  &   \mathtt{\% 8 }   & = &   \pirI{Force}~        (  \mathtt{\% 6 }  )~  \mathtt{e{ 7 } }    \\  \,  &        &   &   \pirI{Return}~       (  \mathtt{\% 8 }  )   \\   \multicolumn{4}{l}{  
                          }
                      \end{array}
                  \]
                 \vskip -0.01em
              } 
The translation of \c
f illustrates the calling convention chosen for \PIR: it requires
environments to be callee-created, \ie callees initialize environments with
arguments.  Accordingly, the \LdArg in \c f loads an argument by position
and  \MkEnv binds it to variable \c b.

We now walk through promise inlining.  First, the callee must be inlined.
Performing inlining at the source level in R is not sound as this would mix
variables defined in different environments.  However, this is not an issue
in \PIR; since environments are modeled explicitly, the inlinee keeps its
local environment as \MkEnv is also inlined.  Therefore, after inlining \c
f, we get:
\vskip -1em
 {\ttfamily\small
                 \vskip -0.5em
                  \[
                      \begin{array}{rlll}
                          \multicolumn{4}{l}{
                               } \\   &   \mathtt{\% 2 }   & = &   \pirI{MkArg}~    (  {\sf pr0 }  ,~ \mathsf{G} )   \\  \,  & \multicolumn{3}{l}{\texttt{\#   {\sf inlinee }   \, } } \\  \,  &   \mathtt{e{ 7 } }   & = &   \pirI{MkEnv}~  (  {\sf b }   \ottsym{=}   \mathtt{\% 2 }  ~  :  \mathsf{G}  )   \\  \,  &   \mathtt{\% 8 }   & = &   \pirI{Force}~        (  \mathtt{\% 2 }  )~  \mathtt{e{ 7 } }    \\   \multicolumn{4}{l}{  
                          }
                      \end{array}
                  \]
                 \vskip -0.01em
              } 

\noindent
The next step is to elide the promise by inlining it where it is evaluated.
We identify the \Force instruction which dominates all other uses of a \MkArg
instruction.  If such a dominating \Force exists, it follows that the
promise must be evaluated at that position.  We inline \c{pr0} to replace
$ \mathtt{\% 8 } $:

\vskip -1em
 {\ttfamily\small
                 \vskip -0.5em
                  \[
                      \begin{array}{rlll}
                          \multicolumn{4}{l}{
                               } \\   &   \mathtt{\% 2 }   & = &   \pirI{MkArg}~    (  {\sf pr0 }  ,~ \mathsf{G} )   \\  \,  & \multicolumn{3}{l}{\texttt{\#   {\sf inlinee }   \, } } \\  \,  &   \mathtt{e{ 6 } }   & = &   \pirI{MkEnv}~  (  {\sf b }   \ottsym{=}   \mathtt{\% 2 }  ~  :  \mathsf{G}  )   \\  \,  & \multicolumn{3}{l}{\texttt{\#   {\sf inlined }  \,  \kern-0.68em   {\sf promise }   } } \\  \,  &   \mathtt{\% 4 }   & = &   \pirI{LdVar}~        (  {\sf x }  ,~ \mathsf{G} )   \\  \,  &   \mathtt{\% 5 }   & = &   \pirI{Force}~        (  \mathtt{\% 4 }  )~ \mathsf{G}   \\   \multicolumn{4}{l}{  
                          }
                      \end{array}
                  \]
                 \vskip -0.01em
              } 

\noindent
We have succeeded in tracking a variable captured by a promise through a
call and evaluation of that promise.  
The $ \mathtt{\% 2 } $ and $e6$ instructions are dead code and can be removed, leaving
only the load and force of \c x.

\bigskip

\section{\PIR in Depth}
\label{sec:pir}

This section describes \PIR in detail.  The IR is introduced in
\autoref{sec:pir-def}.  
Scope resolution, the analysis that tracks R
variables, is presented in \autoref{sec:scope-resolution}.  
Analysis precision is discussed in \autoref{sec:mitigations}. 
A technique to delay the creation of environments is presented in
\autoref{sec:delay-env}.  
Lastly, promise inlining, which builds upon scope resolution, is presented in
\autoref{sec:promise-inline}.

\subsection{Syntax and Semantics}
\label{sec:pir-def}

\begin{figure}\small
\vskip 15pt

\begin{tabular}{l l l}
Function &
  $F ::= \mathit{id} : \ottnt{V}^*$ & labeled list of versions \\
Version &
  $V ::= A~:~\ottnt{C} ~ \, \, \ottnt{P}^*$ & assumptions, body, and promises\\
Promise &
  $P ::= \mathit{id} : \ottnt{C}$ & labeled piece of code \\
Code &
  $C ::= \ottnt{B}^*$ & list of basic blocks\\
Basic Block &
  $B ::= \mathit{L} : \ottnt{st}^*$ & labeled, with a list of statements \\
Label &
  $L ::=  \mathsf{BB}_ \mathit{n} $ & basic block number $n$ \\
\end{tabular}
\caption{Programs}\label{fig:syntax-F}

\medskip

\begin{tabular}{l c l l}
  $v$   & ::= & &\\
        & | & $v_e$ & environment \\
        & | & $v_p$ & promise\\
        & | & $v_c$ & closure \\
        & | & $c$   & constant\\
  $v_e$ & ::= & &\\
        & | & $(x \mapsto v)^* : v_e$ & variables + enclosing env.\\
  $v_p$ & ::= & &\\
        & | & $<\!C, v_e, \_\!>$ & unevaluated promise\\
        & | & $<\!C, v_e, v\!>$  & evaluated promise\\
  $v_c$ & ::= & &\\
        & | & $<\!F, v_e\!>$     & closure \\
\end{tabular}
\caption{Values}\label{fig:syntax-val}

\medskip

\grammartabularSTY{%
  \ottinstr{}\\
  \ottst{}\\
  \ottargOrEnv{}\\
  \ottBinop{}\\
  \ottlit{}
 }
\caption{Instructions}\label{fig:syntax-i}\end{figure}

\autoref{fig:syntax-F} shows the structure of programs.  As
in sourir~\cite{popl18}, each function is versioned. The versions are compiled with
different assumptions ($A$) and have different levels of optimization
applied. Assumptions are predicates that may hold only for some executions
of a function, \eg that arguments are already evaluated.  
\R takes care of calling versions for which all assumptions are satisfied. 
A program is thus a set of functions, each with one or more versions with a
function body and the promises it creates. 
Promise and function bodies are sets of basic blocks.  
Functions, promises, and basic blocks are labeled by names.  
All labels $(\mathit{id})$ are unique.  
Promises and functions in \autoref{fig:syntax-F} should not be confused with
values that represent closures and promises; those are shown in
\autoref{fig:syntax-val}.  
A closure is a pair with a function and its environment, while a promise value
is a triple with code, its environment, and a result. An environment is a
sequence of bindings from variables to values.  

\autoref{fig:syntax-i} shows the remainder of the \PIR grammar.  
\PIR is in SSA form: each statement ($\ottnt{st}$) is constructed such that its
result is assigned to a unique register. While there is only one kind of
register in \PIR, to help readability, our convention is to use 
$(\mathtt{e}\mathit{n})$ for registers that hold environments (or environment
literals) and $(\mathtt{\%}\mathit{n})$ for everything else.  
\PIR has instructions for the following operations: performing arithmetic;
branching; deoptimizing a function; applying a closure; jumping to a basic
block; loading arguments, constants, functions, and variables; creating
promises, environments, and closures; forcing a promise; phi merges; returning
values; and storing variables. 
Most of the instructions are unsurprising (and some have been elided for
brevity). We focus our explanation on \MkEnv, \MkArg, and \Force.

\paragraph{MkEnv.} This instruction takes initial variables and a 
parent environment as arguments:
\[
   \pirI{MkEnv}~  ( {(\mathit{x}=\mathit{a})}^{*} ~ : \mathit{env} ) 
\]
\noindent The resulting environment contains the bindings ${(\mathit{x}=\mathit{a})}^{*}$ and
is scoped inside $\mathit{env}$.  By default functions start out with an
environment that contains all their declared arguments. Thus, a function
defined at the top level with an argument called \c{a} has the following
body:

 {\ttfamily\small
                 \vskip -0.5em
                  \[
                      \begin{array}{rlll}
                           &   \mathtt{\% 0 }   & = &   \pirI{LdArg}~        ( 0 )   \\  \,  &   \mathtt{e{ 1 } }   & = &   \pirI{MkEnv}~  (  {\sf a }   \ottsym{=}   \mathtt{\% 0 }  ~  :  \mathsf{G}  )   \\  \,  &        &   & ... \\  
                      \end{array}
                  \]
                 \vskip -0.01em
              } 

\noindent 
Variables can be added or updated with \StVar and read with \LdVar.  The
latter returns the value of the first binding for the variable in the stack
of environments. That value can be a promise and may or may not be
evaluated. When searching for a function, \LdFun is used instead.  The
instruction evaluates promises and skips over non-function bindings.  One
optimization converts \LdFun into \LdVar when possible.

\paragraph{MkArg.} This instruction creates a promise from an expression 
in the source program and an environment:
\[
   \pirI{MkArg}~    ( \mathit{id} ,~ \mathit{env} ) 
\]
\noindent
The instruction is mainly used to create promises for function arguments.  A
call such as \c{f(a+b)} translates to a load of a function \c f, the
creation of a promise with body \c{p1} and a call:

\begin{minipage}{\columnwidth}
 {\ttfamily\small
                 \vskip -0.5em
                  \[
                      \begin{array}{rlll}
                          \multicolumn{4}{l}{
                               } \\   &   \mathtt{\% 0 }   & = &   \pirI{LdFun}~        (  {\sf f }  ,~  \mathtt{e}\mathit{n}  )   \\  \,  &   \mathtt{\% 1 }   & = &   \pirI{MkArg}~    (  {\sf p1 }  ,~  \mathtt{e}\mathit{n}  )   \\  \,  &   \mathtt{\% 2 }   & = &   \pirI{Call}~    \mathtt{\% 0 }  ~(  \mathtt{\% 1 }  )~  \mathtt{e}\mathit{n}    \\   \multicolumn{4}{l}{  \,  \mathsf{  {\sf p1 }  } } \\   &   \mathtt{\% 0 }   & = &   \pirI{LdVar}~        (  {\sf a }  ,~  \mathtt{e}\mathit{n}  )   \\  \,  &   \mathtt{\% 1 }   & = &   \pirI{LdVar}~        (  {\sf b }  ,~  \mathtt{e}\mathit{n}  )   \\  \,  &   \mathtt{\% 2 }   & = &   \pirI{Force}~        (  \mathtt{\% 0 }  )~  \mathtt{e}\mathit{n}    \\  \,  &   \mathtt{\% 3 }   & = &   \pirI{Force}~        (  \mathtt{\% 1 }  )~  \mathtt{e}\mathit{n}    \\  \,  &   \mathtt{\% 4 }   & = &    \pirI{Add}  ~    (  \mathtt{\% 2 }  ,~  \mathtt{\% 3 }  )~  \mathtt{e}\mathit{n}    \\  \,  &        &   &   \pirI{Return}~       (  \mathtt{\% 4 }  )   \\   \multicolumn{4}{l}{  
                          }
                      \end{array}
                  \]
                 \vskip -0.01em
              } 
\vskip -0.5em
\end{minipage}

\noindent 
The body of the promise contains two reads for \c a and \c b whose results
get forced, a binary addition, and a return. The code is known
statically, while the environment in which it is evaluated is a runtime
value.

\paragraph{Force.}
This instruction takes a promise as input, evaluates it (recursively if
needed), and returns its value:
\[
   \pirI{Force}~        ( \mathit{a} )~ \mathit{env} 
\]
\noindent
Note that $\mathit{env}$ is a synthetic argument that is not needed for
evaluation but describes a data dependency. The promise could access
the current environment using reflective operations. If
$\mathit{a}$ is not a promise then $\mathit{a}$ is returned intact.  If $\mathit{a} =
<\!C,v_e,\_\!>$, then $C$ is evaluated in $v_e$, and the result is stored in the
data structure and returned. Otherwise, if $\mathit{a} = <\!C, v_e, v\!>$, then
$v$ is returned.

\paragraph{Typed Instructions.}
\PIR instructions are typed, which allows more precise register types.  The
types include environments, vectors, scalars, closures, lists, etc.  The type system
also distinguishes between values and both evaluated and unevaluated
promises.  We omit additional details as the types are not relevant for the
optimizations presented in this paper.

\subsection{Scope Resolution}\label{sec:scope-resolution}

Scope resolution is an abstract interpretation over stores.  The
transformation draws inspiration from the \emph{mem2reg} pass in
\emph{LLVM}. \R first compiles variables to environment loads and stores and
later lowers them to registers.  The domain $s$ of the analysis consists of
sequences of abstract environments. Assume that we have environments $ \mathtt{e{ 1 } } , \dots,  \mathtt{e}\mathit{n} $ and variables $\mathit{x}_{{\mathrm{1}}} \dots
\mathit{x}_{\ottmv{m}}$. Then, an abstract state $s$ is a $n * m$ vector of sets of
locations.  A location is either a program point $l$ or $\epsilon$ if the
variable is undefined. We write \get sij to denote the abstract value of
variable $\mathit{x}_{\ottmv{j}}$ in the environment accessed through register $ \mathtt{e}\mathit{i} $.
The value $\top$ denotes the set of all locations---it represents the case
where we do not know anything about a particular variable. The bottom value
is represented by a vector where each element is the empty set.  The
analysis is defined by a transition function over statements and a merge
function over states.

\paragraph{Transition Function.} The transition function takes three
arguments: a program point $l$, a statement $\ottnt{st}$, and an abstract state
$s$. The result is a new abstract state $s'$.  We discuss the three
interesting cases.  Let $\ottnt{st}$ be the creation of a new environment stored
in register $ \mathtt{e}\mathit{i} $ with some values for variables $\mathit{x}_{{\mathrm{1}}}, \dots,
\mathit{x}_{\ottmv{j}}$:
\[
   \mathtt{e}\mathit{i}  =  \pirI{MkEnv}~  ( \mathit{x}_{{\mathrm{1}}}  \ottsym{=}  \mathit{a}_{{\mathrm{1}}}  \ottsym{,} \, ... \, \ottsym{,}  \mathit{x}_{\ottmv{j}}  \ottsym{=}  \mathit{a}_{\ottmv{j}} ~  :   \mathtt{e}\mathit{k}   ) 
\]
\noindent
Then, the resulting state $s'$ is initialized with location $l$ for
variables $\mathit{x}_{{\mathrm{1}}}, \dots, \mathit{x}_{\ottmv{j}}$ in the environment $ \mathtt{e}\mathit{i} $. Other
variables in that environment are set to $\epsilon$ to denote that they are
undefined.
\[
  {\get{(s')}{p}{q}}= \begin{cases}
    \{l\} \quad        & p = i , \, \mathit{x}_{\ottmv{q}} \in \mathit{x}_{{\mathrm{1}}}, \dots, \mathit{x}_{\ottmv{j}} \\
    \{\epsilon\} \quad & p = i , \, \mathit{x}_{\ottmv{q}} \notin \mathit{x}_{{\mathrm{1}}}, \dots, \mathit{x}_{\ottmv{j}} \\
    \get{s}{p}{q} \quad & \text{otherwise}
  \end{cases}
\]
\noindent
For the second interesting case let $\ottnt{st}$ be the store instruction which
defines or updates variable $\mathit{x}_{\ottmv{i}}$ to a value held in register $ \mathtt{\%}\mathit{j} $
for environment $ \mathtt{e}\mathit{k} $:
\[
   \pirI{StVar}~     ( \mathit{x}_{\ottmv{i}} ,~  \mathtt{\%}\mathit{j}  ,~  \mathtt{e}\mathit{k}  ) 
\]
\noindent
This operation simply overwrites the state for that variable $\mathit{x}_{\ottmv{i}}$ with the current
location.
\[
  {\get{(s')}{p}{q}}= \begin{cases}
    \{l\} \quad        & p = k, \, \mathit{x}_{\ottmv{q}} = \mathit{x}_{\ottmv{i}} \\
    \get{s}{p}{q} \quad & \text{otherwise}
  \end{cases}
\]
\noindent
The last case we describe is when an instruction taints the environment, \ie any
instruction that may perform reflective manipulation; this includes \Call,
\Force, and \LdFun.
For example, let $\ottnt{st}$ be a call instruction:
\[
   \pirI{Call}~   \mathit{a}_{{\mathrm{0}}} ~( \mathit{a}_{{\mathrm{1}}}  \ottsym{,} \, .. \, \ottsym{,}  \mathit{a}_{\ottmv{n}} )~  \mathtt{e}\mathit{k}  
\]
\noindent
To be safe, the defined parts of the abstract state are set to $\top$, \ie
we know nothing after this point.
\[
  {\get{(s')}{p}{q}}= \begin{cases}
    \top \quad        & \get{s}{p}{q} \ne \emptyset \\
    \emptyset \quad & \text{otherwise}
  \end{cases}
\]
\noindent
We can improve precision by tracking parent relations between environments
to avoid tainting them all.  Also, the analysis can be extended to be
interprocedural across \Call or \Force instructions.  The state of a scope
resolution in progress can be queried to resolve the target of a \Call or \Force instruction.  Other
mitigations to avoid tainting the state, such as speculative stub
environments or special treatment for non-reflective promises, are discussed
in \autoref{sec:mitigations}.

\paragraph{Merge.}
States are merged at control-flow joins. The merge operation is pointwise set
union.

\paragraph{Transformation.}
Scope resolution computes the reachable stores.
Based on its results, some \LdVar instructions can be removed.
Given a load instruction $ \mathtt{\%}\mathit{i}  =  \pirI{LdVar}~        ( \mathit{x}_{\ottmv{j}} ,~  \mathtt{e}\mathit{k}  ) $ with an abstract
state $s$, there are three possible cases. First, when $\get skj = \{l\}$,
\ie the only observable modification to $\mathit{x}_{\ottmv{j}}$ is by the instruction at
$l$, we can simply replace $ \mathtt{\%}\mathit{i} $ with the register stored by the
instruction at that location.  The second case is $\get skj = \{l_1, \dots,
l_n\}$, \ie depending on the flow of control any one of the $n$ instructions
could have caused the last store.  We use an SSA construction
algorithm~\cite{cyt91} to combine all stored registers in a phi congruence
class.  We replace $ \mathtt{\%}\mathit{i} $ with the \Phi instruction produced by
the SSA construction. Finally, the third case occurs if $\get skj = \top$ or
$\epsilon \in \get skj$, \ie the load cannot be resolved and no
optimization is applied.

\paragraph{Example.}

We conclude with an annotated example of a load that has two flow-dependent
dominating stores:

\begin{lstlisting}
  function () {
    if (...)  x <- 1 
    else      x <- 2
    x
  }
\end{lstlisting}

\noindent The translation starts by creating an empty environment.  After some
branching condition either \c{1} or \c{2} is stored in \c{x}.  Finally, the
value of \c{x} is loaded, forced, and returned.

 {\ttfamily\small
                 \vskip -0.5em
                  \[
                      \begin{array}{rlll}
                           ~~  \mathsf{BB}_ 0  :   &   \mathtt{e{ 1 } }   & = &   \pirI{MkEnv}~  ( \, ~  :  \mathsf{G}  )   \\  \,  &   \mathtt{\% 2 }   & = & ... \\  \,  &        &   &   \pirI{Branch}~  (  \mathtt{\% 2 }  ,~  \mathsf{BB}_ 1  ,~  \mathsf{BB}_ 2  )   \\   \,  ~~  \mathsf{BB}_ 1  :   &   \mathtt{\% 4 }   & = &   \pirI{LdConst}~        [  1  ]~ 1    \\  \,  &        &   &   \pirI{StVar}~     (  {\sf x }  ,~  \mathtt{\% 4 }  ,~  \mathtt{e{ 1 } }  )   \\  \,  &        &   &   \pirI{Branch}~    \mathsf{BB}_ 3    \\   \,  ~~  \mathsf{BB}_ 2  :   &   \mathtt{\% 7 }   & = &   \pirI{LdConst}~        [  1  ]~ 2    \\  \,  &        &   &   \pirI{StVar}~     (  {\sf x }  ,~  \mathtt{\% 7 }  ,~  \mathtt{e{ 1 } }  )   \\  \,  &        &   &   \pirI{Branch}~    \mathsf{BB}_ 3    \\   \,  ~~  \mathsf{BB}_ 3  :   &   \mathtt{\% 10 }   & = &   \pirI{LdVar}~        (  {\sf x }  ,~  \mathtt{e{ 1 } }  )   \\  \,  &   \mathtt{\% 11 }   & = &   \pirI{Force}~        (  \mathtt{\% 10 }  )~  \mathtt{e{ 1 } }    \\  \,  &        &   &   \pirI{Return}~       (  \mathtt{\% 11 }  )   \\   
                      \end{array}
                  \]
                 \vskip -0.01em
              } 

\noindent
This function has one environment ($ \mathtt{e{ 1 } } $) and one variable (\c x), thus it
is represented as vector of length one, starting empty $\langle\{\}\rangle$.
The scope analysis derives an abstract state $\langle\{5,8\}\rangle$ (where
5 and 8 are the locations of both stores).  Therefore we place a
\Phi instruction in $ \mathsf{BB}_ 3 $ to join those two writes.  We can
replace the load $ \mathtt{\% 10 } $ with this phi.

 {\ttfamily\small
                 \vskip -0.5em
                  \[
                      \begin{array}{rlll}
                           ~~  \mathsf{BB}_ 0  :   &   \mathtt{\% 1 }   & = & ... \\  \,  &        &   &   \pirI{Branch}~  (  \mathtt{\% 2 }  ,~  \mathsf{BB}_ 1  ,~  \mathsf{BB}_ 2  )   \\   \,  ~~  \mathsf{BB}_ 1  :   &   \mathtt{\% 4 }   & = &   \pirI{LdConst}~        [  1  ]~ 1    \\  \,  &        &   &   \pirI{Branch}~    \mathsf{BB}_ 3    \\   \,  ~~  \mathsf{BB}_ 2  :   &   \mathtt{\% 7 }   & = &   \pirI{LdConst}~        [  1  ]~ 2    \\  \,  &        &   &   \pirI{Branch}~    \mathsf{BB}_ 3    \\   \,  ~~  \mathsf{BB}_ 3  :   &   \mathtt{\% 10 }   & = &   \pirI{Phi}~  (   \mathsf{BB}_ 1   :   \mathtt{\% 4 }    \ottsym{,}    \mathsf{BB}_ 2   :   \mathtt{\% 7 }   )   \\  \,  &        &   &   \pirI{Return}~       (  \mathtt{\% 10 }  )   \\   
                      \end{array}
                  \]
                 \vskip -0.01em
              } 

\noindent
Since the load is statically resolved, dead store elimination is able to
remove both \StVar instructions. Combined with an escape analysis, the
environment is also elided.  The \Force instruction is removed, since we
know that the value is either of the two constants and not a promise.

\subsection{Improving Precision}\label{sec:mitigations}
The problem with the analysis presented so far is that, because of R's semantics,
any instruction that evaluates potentially effectful code taints the
abstract environment.  
There are two kinds of non-local effects: callees may
affect functions upwards the call stack by accessing the caller's
environment, and callers pass promises that may affect functions downwards
the call stack, when those functions force the promises. 
To make scope resolution useful in practice, the impact of these
non-local effects should be somewhat mitigated. For instance, we
rely on inlining to reduce the number of \Call and \Force instructions.
Below we explain a special treatment for non-reflective promises and 
a speculative optimization assuming calls do not change the environment.

\paragraph{Contextual Assumptions.}

We leverage the fact that \PIR functions can have multiple versions
optimized under different assumptions to treat some promises
specially by making contextual assumptions.  
For instance, if all arguments to a call are values, it is safe to invoke a
version of the function that ignores the dangers of promise evaluation.  
Further specializations are possible for pure promises, or non-reflective ones.
This specialization trades performance for code size, since functions must be
compiled multiple times.
\R dynamically picks the optimal version of a function to invoke.  

\paragraph{Stub Environments.}

If an environment is locally resolved, but could be tainted during a call using reflection, then
we speculatively elide that environment and replace it by a stub.  At
runtime a stub environment has the same structure as a normal environment
shown in \autoref{fig:syntax-val}, but a more compact representation, since
it does not need to support updates.  If the stub environment is modified
then it is transparently converted into a full environment.  In \PIR, stub
environments are created by a structurally identical variant of the \MkEnv
instruction.  After a call we check if the stub was materialized, in
which case we deoptimize the current function.  Consequently, 
analyses on \PIR can assume stub environments to not experience any
non-local modifications.

\subsection{Delaying Environments}\label{sec:delay-env}
\R has a deoptimization mechanism that transfers control back to the unoptimized
version of a function.  
A deoptimization point includes the following instructions:

 {\ttfamily\small
                 \vskip -0.5em
                  \[
                      \begin{array}{rlll}
                           ~~  \mathsf{BB}_ 0  :   &   \mathtt{e{ 1 } }   & = &   \pirI{MkEnv}~  (  {\sf foo }   \ottsym{=}   \mathtt{\%}\mathit{i}  ~  :   \mathtt{e{ 0 } }   )   \\  \,  &        &   & ... \\  \,  &   \mathtt{\% 2 }   & = & ... \\  \,  &        &   &   \pirI{Branch}~  (  \mathtt{\% 2 }  ,~  \mathsf{BB}_ 2  ,~  \mathsf{BB}_ 1  )   \\   \,  ~~  \mathsf{BB}_ 1  :   &        &   &   \pirI{Deopt}~(  {\sf baselinePc }  ,~  \mathtt{\%}\mathit{1}, \dots, \mathtt{\%}\mathit{n}  ,~  \mathtt{e{ 1 } }  )   \\   
                      \end{array}
                  \]
                 \vskip -0.01em
              } 

\noindent
An assumption ($ \mathtt{\% 2 } $) is checked by the \Branch instruction.  Any boolean
instruction can be used as an assumption.  In case it holds, we continue to
$ \mathsf{BB}_ 2 $, otherwise the deoptimization branch $ \mathsf{BB}_ 1 $ is entered.
The unconditional deoptimization instruction, \Deopt, contains all the
metadata needed to transfer control back to the baseline version of the
function.  The arguments $ \mathtt{\%}\mathit{1}, \dots, \mathtt{\%}\mathit{n} $ are the values that the target code
expects on the operand stack.  
Finally, since the baseline version in \R requires the
environment of a function to be always present, the \Deopt
instruction requires it to be present.

To avoid creating an environment at runtime whenever a compiled function
performs any kind of speculative optimization, creation of environments should
be delayed as much as possible.
Optimizations are allowed to move \MkEnv instructions into
branches and even over writes to that environment. When that happens, the
\StVar is removed and the value is added to the initialization list.
When an environment is used by multiple deoptimization points, then this is not
sufficient, since each deoptimization branch will require the environment in a different state.

Partial escape analysis~\cite{Stadler14} intends to delay an allocation to
only those branches where the object escapes.
Similarly, in \PIR we aim to materialize an up-to-date environment in each
deoptimization branch, allowing us to elide the environment in the main path.
This requires replaying stores between the original environment creation and the
\Deopt instruction.
We use the output of scope analysis to determine the state of the environment in the
deoptimization branch.
A sufficient condition is that at the \Deopt instruction none of the variables
in the abstract state $s$ (see sec. 4.2) is $\top$.

Assume there is an $ \pirI{StVar}~     (  {\sf bar }  ,~  \mathtt{\%}\mathit{j}  ,~  \mathtt{e{ 1 } }  ) $ instruction between \MkEnv and \Deopt in the previous example.
We now proceed to assemble a synthetic environment to replace $ \mathtt{e{ 1 } } $ in the deoptimization branch.
In this case the abstract environment $ \mathtt{e{ 1 } } $ according to scope resolution is such that \c{foo} is defined by the \MkEnv instruction and \c{bar} is defined by the \StVar instruction.
In \autoref{sec:scope-resolution} we presented our technique to replace a \LdVar instruction with a \PIR register.
We now reuse the same technique to capture the current values of \c{foo} and \c{bar} as registers and then include them in a fresh \MkEnv instruction:
 {\ttfamily\small
                 \vskip -0.5em
                  \[
                      \begin{array}{rlll}
                           ~~  \mathsf{BB}_ 0  :   &   \mathtt{e{ 1 } }   & = &   \pirI{MkEnv}~  (  {\sf foo }   \ottsym{=}   \mathtt{\%}\mathit{i}  ~  :   \mathtt{e{ 0 } }   )   \\  \,  &        &   & ... \\  \,  &        &   &   \pirI{StVar}~     (  {\sf bar }  ,~  \mathtt{\%}\mathit{j}  ,~  \mathtt{e{ 1 } }  )   \\  \,  &   \mathtt{\% 2 }   & = & ... \\  \,  &        &   &   \pirI{Branch}~  (  \mathtt{\% 2 }  ,~  \mathsf{BB}_ 2  ,~  \mathsf{BB}_ 1  )   \\   \,  ~~  \mathsf{BB}_ 1  :   &   \mathtt{e{ 3 } }   & = &   \pirI{MkEnv}~  (  {\sf foo }   \ottsym{=}   \mathtt{\%}\mathit{i}   \ottsym{,}   {\sf bar }   \ottsym{=}   \mathtt{\%}\mathit{j}  ~  :   \mathtt{e{ 0 } }   )   \\  \,  &        &   &   \pirI{Deopt}~(  {\sf baselinePc }  ,~  \mathtt{\%}\mathit{1}, \dots, \mathtt{\%}\mathit{n}  ,~  \mathtt{e{ 3 } }  )   \\   
                      \end{array}
                  \]
                 \vskip -0.01em
              } 

\noindent
Assuming we are able to materialize a copy of the environment in
every deoptimization branch, it is then possible to remove the original \MkEnv.
This transformation duplicates variables for each deoptimization branch.
They can later be cleaned up using some form of redundancy elimination, such
as global value numbering.

Contrary to replacing $\pirI{LdVar}$s, it is
possible to materialize environments even when the analysis results contain
$\epsilon$.  For those cases a runtime marker is used to indicate the
absence of a binding. \MkEnv will simply skip a particular binding if its
input value is equal to this marker value.

\subsection{Promise Inlining}\label{sec:promise-inline}

The analysis needed for promise inlining is a simple dataflow analysis.  The
values of interest are promises created by \MkArg.  The analysis
uses a lattice for the state of a promise that starts at bottom, $\bot$,
and can either be forced at program point $l$ or leaked ($\triangledown$),
and tops at $\top$.  There is one such state per promise-creating
instruction, thus the abstract state is a vector of length $n$ where $n$ is
the number of \MkArg instructions in the function.  We present the
abstract interpretation by discussing the transition function that takes a
statement and an abstract state, and returns a new abstract state, and the
merge function that combines two states.

\paragraph{Transition Function.}
The abstract state is initialized to $\bot^*$.  There are three interesting
cases.

First, given an instruction $ \pirI{Force}~        (  \mathtt{\%}\mathit{i}  )~  \mathtt{e}\mathit{j}  $ at
location $l$, where $ \mathtt{\%}\mathit{i} $ is a \MkArg instruction, we update the abstract
state of the promise $ \mathtt{\%}\mathit{i} $ as follows: if the state is $\bot$, then it is set
to $l$, indicating that this is the dominating \Force.
If the state is $\triangledown$ the result is $\top$, otherwise it stays unchanged.

Second, given an instruction $ \pirI{MkEnv}~  ( \mathit{x}_{{\mathrm{1}}}  \ottsym{=}  \mathit{a}_{{\mathrm{1}}}  \ottsym{,} \, ... \, \ottsym{,}  \mathit{x}_{\ottmv{j}}  \ottsym{=}  \mathit{a}_{\ottmv{j}} ~  :   \mathtt{e}\mathit{p}   ) $, for any input $\mathit{a}_{{\mathrm{1}}}, \dots,
\mathit{a}_{\ottmv{j}}$ that refers to a promise, the state of that promise is set to
leaked ($\triangledown$) if it is $\bot$, otherwise it stays unchanged. 

Third, given any instruction which could evaluate promises, such
as \Call, \Force, or \LdFun, all escaped ($\triangledown$) promises are updated to $\top$.

Therefore, promises used first in a \MkEnv and then in a \Force instructions will end up at
$\top$ and not be inlined.  If we were to inline such a promise it would
cause the result slots of the promise in the environment to be out of sync
with the result of the inlined expression.  It is possible to support some
of those edge cases with an instruction to update the result slot of a promise.

\paragraph{Merge.}
When merging abstract states, identical states remain and disagreeing states
become $\top$.  The latter can happen for example in

\begin{lstlisting}
  function(a) {
    if(...)  a
    a
  }
\end{lstlisting}

\noindent where \c{a} is forced depends on a condition; it could be either
line 1 or 2.  While it would be possible to track those cases more
accurately, we did not need it in practice yet.

\paragraph{Code Transformation.}

The promise inlining pass uses the analysis to inject promises at their
dominating force instruction.  As a precondition, we need a \MkArg and the
corresponding \Force instruction to be in the same function.  This only
happens after inlining, since initially creation and evaluation of promises
is always separated by a call.  The promise inliner will
inline the promise body at the location of the dominating force, update all
uses with the result of the inlinee, and remove both the \MkArg and the
\Force.

If this promise originates from a \LdArg instruction, then the
promise originates from an argument passed to the current function.  We do
not know its code and therefore cannot inline it.  On the other hand we can
still replace all uses of the dominated \Force instruction with the
dominating \Force instruction.

\paragraph{Example.}

\begin{figure}\begin{lstlisting}
  g <- function() {
    a <- 1
    f <- function(b) b+a
    f(2)
  }
\end{lstlisting}\vskip -1em
\caption{An example with promises to be inlined.}\label{fig:example-prom-inline}

\medskip

 {\ttfamily\small
                 \vskip -0.5em
                  \[
                      \begin{array}{rlll}
                          \multicolumn{4}{l}{
                               \mathsf{  {\sf g }  } } \\   &   \mathtt{\% 7 }   & = &   \pirI{LdConst}~        [  1  ]~ 1    \\  \,  &   \mathtt{e{ 8 } }   & = &   \pirI{MkEnv}~  (  {\sf a }   \ottsym{=}   \mathtt{\% 7 }  ~  :  \mathsf{G}  )   \\  \,  &   \mathtt{\% 9 }   & = &   \pirI{MkClosure}~(  {\sf f }  ,~  \mathtt{e{ 8 } }  )   \\  \,  &   \mathtt{\% 10 }   & = &   \pirI{MkArg}~    (  {\sf pr0 }  ,~  \mathtt{e{ 8 } }  )   \\  \,  &   \mathtt{\% 11 }   & = &   \pirI{Call}~    \mathtt{\% 9 }  ~(  \mathtt{\% 10 }  )~  \mathtt{e{ 8 } }    \\  \,  &   \mathtt{\% 12 }   & = &   \pirI{Force}~        (  \mathtt{\% 11 }  )~  \mathtt{e{ 8 } }    \\  \,  &        &   &   \pirI{Return}~       (  \mathtt{\% 12 }  )   \\   \multicolumn{4}{l}{  \,  \mathsf{  {\sf pr0 }  } } \\   &   \mathtt{\% 13 }   & = &   \pirI{LdConst}~        [  1  ]~ 2    \\  \,  &        &   &   \pirI{Return}~       (  \mathtt{\% 13 }  )   \\   \multicolumn{4}{l}{  \,  \mathsf{  {\sf f }  } } \\   &   \mathtt{\% 1 }   & = &   \pirI{LdArg}~        ( 0 )   \\  \,  &   \mathtt{e{ 2 } }   & = &   \pirI{MkEnv}~  (  {\sf b }   \ottsym{=}   \mathtt{\% 1 }  ~  :  \mathsf{O}  )   \\  \,  &   \mathtt{\% 3 }   & = &   \pirI{Force}~        (  \mathtt{\% 1 }  )~  \mathtt{e{ 2 } }    \\  \,  &   \mathtt{\% 4 }   & = &   \pirI{LdVar}~        (  {\sf a }  ,~  \mathtt{e{ 2 } }  )   \\  \,  &   \mathtt{\% 5 }   & = &   \pirI{Force}~        (  \mathtt{\% 4 }  )~  \mathtt{e{ 2 } }    \\  \,  &   \mathtt{\% 6 }   & = &    \pirI{Add}  ~    (  \mathtt{\% 3 }  ,~  \mathtt{\% 5 }  )~  \mathtt{e{ 2 } }    \\  \,  &        &   &   \pirI{Return}~       (  \mathtt{\% 6 }  )   \\   \multicolumn{4}{l}{  
                          }
                      \end{array}
                  \]
                 \vskip -0.01em
              } 
\vskip -2em
\caption{\PIR translation of the function from \autoref{fig:example-prom-inline}.}\label{fig:example-prom-inline2}

\medskip
 {\ttfamily\small
                 \vskip -0.5em
                  \[
                      \begin{array}{rlll}
                          \multicolumn{4}{l}{
                               \mathsf{  {\sf g }  } } \\   &   \mathtt{\% 7 }   & = &   \pirI{LdConst}~        [  1  ]~ 1    \\  \,  &   \mathtt{e{ 8 } }   & = &   \pirI{MkEnv}~  (  {\sf a }   \ottsym{=}   \mathtt{\% 7 }  ~  :  \mathsf{G}  )   \\  \,  &   \mathtt{\% 10 }   & = &   \pirI{MkArg}~    (  {\sf pr0 }  ,~  \mathtt{e{ 8 } }  )   \\  \,  & \multicolumn{3}{l}{\texttt{\#   {\sf inlinee }  \,  \kern-0.68em   {\sf begins }   } } \\  \,  &   \mathtt{e{ 2 } }   & = &   \pirI{MkEnv}~  (  {\sf b }   \ottsym{=}   \mathtt{\% 10 }  ~  :   \mathtt{e{ 8 } }   )   \\  \,  &   \mathtt{\% 3 }   & = &   \pirI{Force}~        (  \mathtt{\% 10 }  )~  \mathtt{e{ 2 } }    \\  \,  &   \mathtt{\% 4 }   & = &   \pirI{LdVar}~        (  {\sf a }  ,~  \mathtt{e{ 2 } }  )   \\  \,  &   \mathtt{\% 5 }   & = &   \pirI{Force}~        (  \mathtt{\% 4 }  )~  \mathtt{e{ 2 } }    \\  \,  &   \mathtt{\% 6 }   & = &    \pirI{Add}  ~    (  \mathtt{\% 3 }  ,~  \mathtt{\% 5 }  )~  \mathtt{e{ 2 } }    \\  \,  & \multicolumn{3}{l}{\texttt{\#   {\sf inlinee }  \,  \kern-0.68em   {\sf ends }   } } \\  \,  &   \mathtt{\% 12 }   & = &   \pirI{Force}~        (  \mathtt{\% 6 }  )~  \mathtt{e{ 8 } }    \\  \,  &        &   &   \pirI{Return}~       (  \mathtt{\% 12 }  )   \\   \multicolumn{4}{l}{  
                          }
                      \end{array}
                  \]
                 \vskip -0.01em
              } 
\vskip -2em
\caption{After inlining function \c f in \autoref{fig:example-prom-inline2}.}\label{fig:example-prom-inline3}

\medskip
 {\ttfamily\small
                 \vskip -0.5em
                  \[
                      \begin{array}{rlll}
                          \multicolumn{4}{l}{
                               \mathsf{  {\sf g }  } } \\   &   \mathtt{\% 7 }   & = &   \pirI{LdConst}~        [  1  ]~ 1    \\  \,  &   \mathtt{e{ 8 } }   & = &   \pirI{MkEnv}~  (  {\sf a }   \ottsym{=}   \mathtt{\% 7 }  ~  :  \mathsf{G}  )   \\  \,  &   \mathtt{\% 10 }   & = &   \pirI{MkArg}~    (  {\sf pr0 }  ,~  \mathtt{e{ 8 } }  )   \\  \,  & \multicolumn{3}{l}{\texttt{\#   {\sf inlinee }  \,  \kern-0.68em   {\sf begins }   } } \\  \,  &   \mathtt{e{ 2 } }   & = &   \pirI{MkEnv}~  (  {\sf b }   \ottsym{=}   \mathtt{\% 10 }  ~  :   \mathtt{e{ 8 } }   )   \\  \,  & \multicolumn{3}{l}{\texttt{\#   {\sf inlined }  \,  \kern-0.68em   {\sf promise }   \,  \kern-0.68em   {\sf begins }   } } \\  \,  &   \mathtt{\% 13 }   & = &   \pirI{LdConst}~        [  1  ]~ 2    \\  \,  & \multicolumn{3}{l}{\texttt{\#   {\sf inlined }  \,  \kern-0.68em   {\sf promise }   \,  \kern-0.68em   {\sf ends }   } } \\  \,  &   \mathtt{\% 6 }   & = &    \pirI{Add}  ~    (  \mathtt{\% 13 }  ,~  \mathtt{\% 7 }  )~  \mathtt{e{ 2 } }    \\  \,  & \multicolumn{3}{l}{\texttt{\#   {\sf inlinee }  \,  \kern-0.68em   {\sf ends }   } } \\  \,  &        &   &   \pirI{Return}~       (  \mathtt{\% 6 }  )   \\   \multicolumn{4}{l}{  
                          }
                      \end{array}
                  \]
                 \vskip -0.01em
              } 
\vskip -2em
\caption{After inlining promise \c pr0 in \autoref{fig:example-prom-inline3}.}\label{fig:example-prom-inline4}
\end{figure}

We now present an example that combines scope resolution and promise
inlining, shown in \autoref{fig:example-prom-inline}.  An inner closure \c f is
called with \c 2 as an argument.  \c f captures the binding of \c a from its
parent environment.  This translates (after scope resolution) to the \PIR code shown in \autoref{fig:example-prom-inline2}.
The parent environment $ \mathsf{O} $ denotes the environment
supplied by \MkClosure.  Since \c{f} is an inner function, it needs to be
closed over the environment at its definition.

The first step necessary to
get the promise creation and evaluation into the same \PIR function, is to
inline the inner function \c f.  After this transformation we obtain the code in \autoref{fig:example-prom-inline3}.
The open environment $ \mathsf{O} $ is replaced with $ \mathtt{e{ 8 } } $.
And the argument $ \pirI{LdArg}~        ( 0 ) $ of the callee is replaced by the
$ \pirI{MkArg}~    (  {\sf pr0 }  ,~  \mathtt{e{ 8 } }  ) $ instruction of the caller.

Now, we can identify the
dominating \Force instruction at $ \mathtt{\% 3 } $.  Therefore, the promise inliner
replaces the \Force instruction with the body of the promise, yielding the
result in \autoref{fig:example-prom-inline4} (after another scope resolution pass).

Only after these steps finish can traditional compiler
optimizations, such as escape analysis on the environment, dead code
elimination, and constant folding, reduce the code to a single
\LdConst instruction.

\section{Results}\label{sec:results}

In this section, we assess scope resolution and promise inlining as means to
statically resolve bindings and reduce the number of environments and
promises needed by R programs.  To do so, we present three
experiments,%
each designed to answer one of the following questions:

\begin{itemize}
\item[{\bf RQ1}] \emph{What proportion of function definitions do not require an environment after optimizations?}
\item[{\bf RQ2}] \emph{What proportion of function invocations do not require an environment after optimizations?}
\item[{\bf RQ3}] \emph{What is the performance impact of scope resolution and promise inlining on the rest of the optimizations?}
\end{itemize}

\paragraph{Methodology.}
Measurements are gathered using \R.  Our first and second experiments rely
on instrumenting the compiler to record information about code being compiled
and also dynamic counters of events happening at runtime. For the last
experiment, which looks at the impact of optimizations, we selected programs
whose performance was impacted by \R. To gather the measurements, we ran 5
invocations with 15 iterations of each benchmark on an Intel i7-3520M CPU,
stepping 9, microcode version 0x21, a clock pinned at 1.2 GHz, on a Fedora
28 running Linux 5.0.16-100, with SpeedStep and lower C-States disabled.
Our compiler has not been written for speed or optimized, so we discard the
first 5 iterations to amortize compilation time.%
\footnote{The experiments are published in runnable form at
  \url{gitlab.com/rirvm/rir\_experiments/container\_registry}. All reported
  numbers are for revision:\\ \tiny
  dba88e9bc417325a29c91acb088df7fe8109ca39e427e03931114e0715513bfafcd59a267812dcb1.}
We do not provide results
for real-world applications, as our system is not ready to compete with
mature R implementations.

\subsection{Static Environment Reduction}

\begin{figure}[!t]
\fbox{\begin{minipage}{0.45\textwidth} \small\raggedright \textbf{Code
      usage} is \c{check_code_usage_in_packages}, an analysis function
    shipped with GNU R. \textbf{Demos} runs all demos from the base
    packages.  \textbf{Tcltk}, \textbf{Stats}, and \textbf{Utils} run all
    examples included, respectively, in the \c{tcltk}, \c{stats} and
    \c{utils} packages.  \textbf{Pidigits} is from the language shootout
    benchmarks.  \textbf{Mandelbrot} is from the
    \emph{are-we-fast} benchmarks~\citep{Mar16}.
\end{minipage}}
\skip -0.5em
\caption{Description of benchmarks for {\bf RQ1} and {\bf RQ2}.}
\label{fig:example-progs}
\vskip -0.6em
\end{figure}

For {\bf RQ1}, we count the number of {\MkEnv}s in compiled code.  We
distinguish between stubs and standard environments.  Note that we ignore
the environments created in deoptimization branches to allow transferring
from optimized code to the baseline version.  \autoref{fig:example-progs}
lists the programs analyzed. \autoref{fig:static-elision} lists the number
of closures that are compiled (Closures), the percentage of
closures which have environments (Env), the percentage of closures that use
stubs (Stub), and the percentage of closures that have no \MkEnv instruction (No
env).  Stubs are inserted when our analysis determines that an environment
is only accessible through reflection.  Adding Stub and No Env, between 12\%
and 65\% of environments are elided.

\begin{figure}[!h]\small\begin{tabular}{l r r r r}
  Program & Closures & Env & Stub & No env \\ \hline
  Code usage & 971  & 82\% & 1\% & 16\% \\
  Demos      & 381  & 81\% & 16\% &3\% \\
  Tcltk      & 18   & 83\% & 11\% &5\% \\
  Stats      & 871  & 85\% & 13\% &2\% \\
  Utils      & 471  & 88\% & 10\% &2\% \\
  Pidigits   & 139  & 79\% & 4\% & 17\% \\
  Mandelbrot & 14   & 36\% & 29\% &36\% \\
\end{tabular}
\vskip -0.5em
\caption{Ratio of statically elided environments}\label{fig:static-elision}
\end{figure}
\vskip -0.4em

\subsection{Dynamic Environment Reduction}

To answer {\bf RQ2}, we counted {\MkEnv}s executed at runtime with and
without optimizations.  \autoref{fig:dynamic-elision} lists the number of
environments allocated in the non-optimized version (Baseline), the
percentage reduction in allocated environments (Reduction), and the
percentage of the reduction that is due to stubs (Stubbed). For instance,
consider a program where the baseline allocates 100 environments and the optimized version only
50, with 10 stubs. This means we reduced the number of
environments by 50\%, and 20\% of that reduction is achieved by stubbing.
The data suggest that, for our benchmarks, 28\% to 87\% fewer environments
were created.

\begin{figure}[!h]\small
\begin{tabular}{l r r r r}
  Program & Baseline & Reduction & Stubbed \\
  \hline
  Code usage     & 2445996 & 36\% & 2\% \\
  Demos          & 192772  & 28\% & 5\% \\
  Tcltk          & 1271    & 43\% & 3\% \\
  Stats          & 3046614 & 28\% & 7\% \\
  Utils          & 2792534 & 33\% & 2\% \\
  Pidigits       & 9032031 & 31\% & 2\% \\
  Mandelbrot     & 8460641 & 87\% & 8\% \\
\end{tabular}
\vskip -0.5em
\caption{Reduction in environments allocated}\label{fig:dynamic-elision}
\end{figure}

\subsection{Effects on Optimizations}

Finally, we address {\bf RQ3} by providing slowdowns between \R with and
without optimizations. \R performs traditional optimizations, such as
constant folding, dead code elimination, and global value numbering.
We posit that those optimization are helped by scope resolution and promise
inlining.
To test this hypothesis, we measure the impact of disabling those
passes on running time.
We run in four configurations: (1) default, (2)
without promise inlining,
\begin{figure}[b]\begin{centering}
\includegraphics[width=0.47\textwidth]{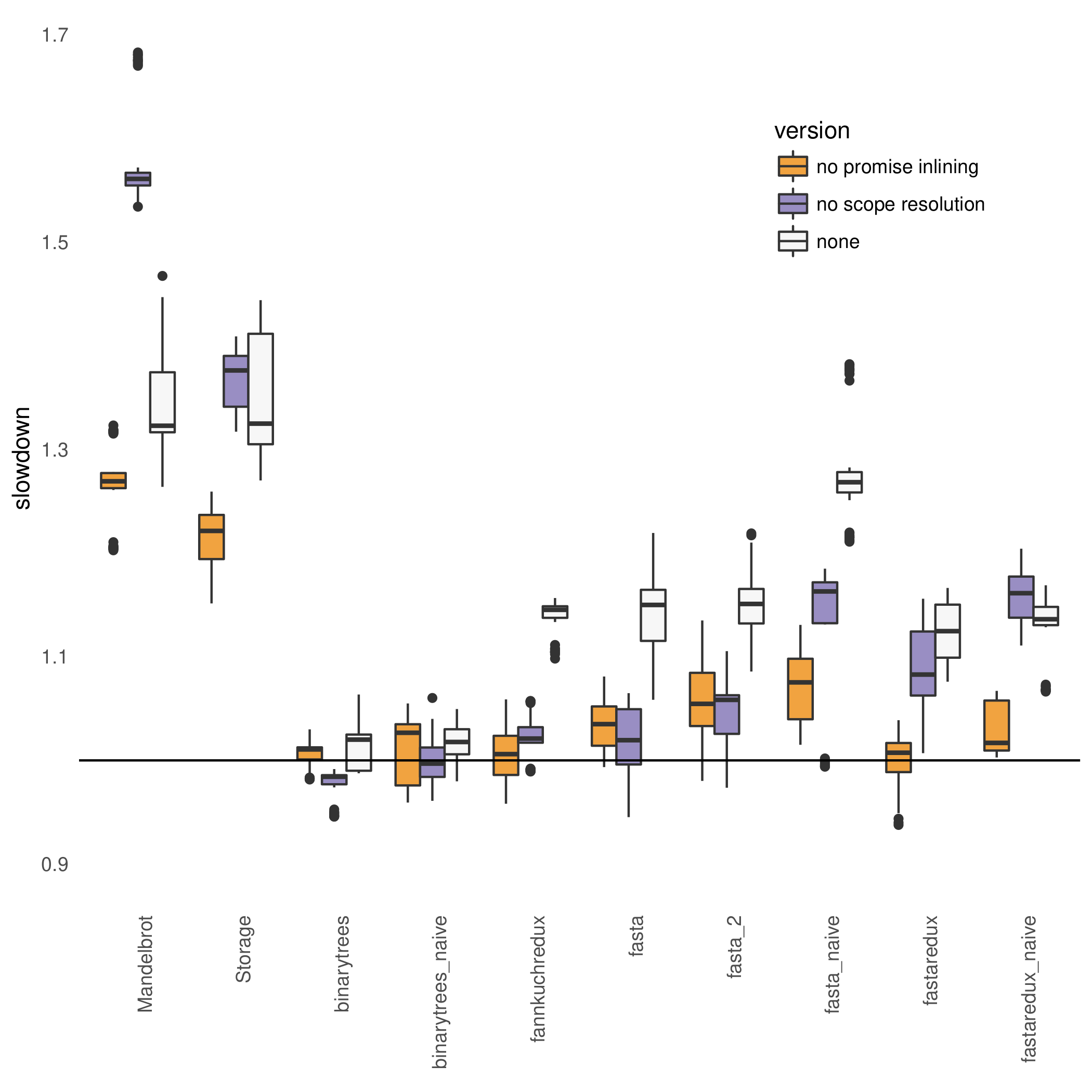}
\caption{Slowdown with promise inlining (orange), scope resolution (purple), or all
optimizations (white) turned off.}\label{fig:perf}\end{centering}
\vskip -1.2mm
\end{figure}
(3) without scope resolution, and (4) with no
optimizations. \autoref{fig:perf} shows slowdowns relative to configuration
(1).  
Disabling all optimizations slows down code by a range from
2\% to 32\%. Much of this slowdown can be attributed to the scope resolution
pass, which when turned off, slows down execution from -1\% to
55\%. Disabling promise inlining has a smaller effect, ranging from a 0\% to
a 35\% slowdown.  Looking at the programs individually, we observe that
\c{Mandelbrot} is surprising in that turning off scope resolution generates
code that is slower than with no optimizations at all.  A reason is that in
this configuration, we still create stub environments and guards, while
{scope resolution} would be the main consumer of this speculation. Here they
only add overheads for the additional deoptimization points.  Disabling
scope resolution contributes significantly to the slowdown in
\c{Fastaredux_naive}.  \c{Storage}, \c{fannkuch_redux}, \c{fasta},
\c{fasta_2}, \c{fasta_naive} all have the expected behavior. As for
\c{binarytrees} and \c{binarytrees_naive}, they show little slowdown in any
configuration.

\section{Conclusion}\label{sec:conclusion}

{\small\begin{quote}
\emph{<<Working on the thing can drive you mad. That's what happened to
  this friend of mine. So he had a lobotomy. Now he's well again.>>}
-- Repo Man
\end{quote}}

\noindent
Designing an intermediate representation for R has been a surprisingly
difficult endeavor.  Our goal was to arrive at a code format that captures
the intricacies of the language while enabling compiler optimizations.  Our
explicit goals were to distinguish between arguments that need lazy
evaluation and ones that do not, to distinguish between variables that are
truly local and can be optimized and variables that must be allocated in
environments and may be exposed through reflection, to allow for elision of
environments when they are known to not be needed. To achieve this, we
designed \PIR, the intermediate representation of the \R compiler. It has
explicit instructions for creating environments, creating promises, and
evaluating promises.  Explicit modeling of constructs that are to be
optimized away is a key design ingredient.  For example, explicit
environments allow functions to be inlined without fully resolving all R
variables upfront.

The challenge presented by R is that it requires solving many problems at
once.  To get rid of laziness, one must track the flow of arguments and
understand where they may be forced.  To track arguments, one has to reason
about environments and how they are manipulated.  To discover if
environments change, one has to analyze promises.  This paper lays out our
current strategy for dealing with this particular mix of dynamic features.

We illustrate the benefits of \PIR with two optimizations, scope resolution
and promise inlining.  Scope resolution statically resolves bindings to
reduce the issue of tracking loads and stores in environments. When
successful, this pass lowers R variables to \PIR registers.  Explicit
creation and evaluation of promises facilitates the inlining of promises.
In combination, these transformations produce code with fewer promises and
first-class environments. Evaluation shows encouraging results: up to 65\%
of functions do not need an R environment, resulting in up to 87\% fewer environments created at runtime.

While \PIR was designed for \R, the design principles generalize to other
implementations of the language. Other languages which have either lazy
evaluation or first-class environments could adopt similar ideas in their
intermediate representations. 

\begin{acks}
We thank the anonymous reviewers and Tom\'a\v{s} Kalibera, for their
insightful comments and suggestions to improve this paper.  This work has
received funding from the \grantsponsor{ONR}{Office of Naval Research
  (ONR)}{} award \grantnum{ONR}{503353}, the \grantsponsor{NSF}{National
  Science Foundation}{} awards \grantnum{NSF}{1544542} and
\grantnum{NSF}{1618732}, the \grantsponsor{BC}{Czech Ministry of Education,
  Youth and Sports from the Czech Operational Programme Research,
  Development, and Education}{}, under grant agreement No.
\grantnum{BC}{CZ.02.1.01/0.0/0.0/15\_003/0000421}, and the
\grantsponsor{ELE}{European Research Council (ERC) under the European
  Union's Horizon 2020 research and innovation programme}{}, under grant
agreement No.  \grantnum{ELE}{695412}.
\end{acks}

\bibliography{../bib/pir.bib,../bib/jv.bib}

\end{document}

%% file: pir_.tex
\newcommand{\ottdrule}[4][]{{\displaystyle\frac{\begin{array}{l}#2\end{array}}{#3}\quad\ottdrulename{#4}}}
\newcommand{\ottusedrule}[1]{\[#1\]}
\newcommand{\ottpremise}[1]{ #1 \\}
\newenvironment{ottdefnblock}[3][]{ \framebox{\mbox{#2}} \quad #3 \\[0pt]}{}
\newenvironment{ottfundefnblock}[3][]{ \framebox{\mbox{#2}} \quad #3 \\[0pt]\begin{displaymath}\begin{array}{l}}{\end{array}\end{displaymath}}
\newcommand{\ottfunclause}[2]{ #1 \equiv #2 \\}
\newcommand{\ottnt}[1]{\mathit{#1}}
\newcommand{\ottmv}[1]{\mathit{#1}}
\newcommand{\ottkw}[1]{\mathbf{#1}}
\newcommand{\ottsym}[1]{#1}
\newcommand{\ottcom}[1]{\text{#1}}
\newcommand{\ottdrulename}[1]{\textsc{#1}}
\newcommand{\ottcomplu}[5]{\overline{#1}^{\,#2\in #3 #4 #5}}
\newcommand{\ottcompu}[3]{\overline{#1}^{\,#2<#3}}
\newcommand{\ottcomp}[2]{\overline{#1}^{\,#2}}
\newcommand{\ottgrammartabular}[1]{\begin{supertabular}{llcllllll}#1\end{supertabular}}
\newcommand{\ottmetavartabular}[1]{\begin{supertabular}{ll}#1\end{supertabular}}
\newcommand{\ottrulehead}[3]{$#1$ & & $#2$ & & & \multicolumn{2}{l}{#3}}
\newcommand{\ottprodline}[6]{& & $#1$ & $#2$ & $#3 #4$ & $#5$ & $#6$}
\newcommand{\ottfirstprodline}[6]{\ottprodline{#1}{#2}{#3}{#4}{#5}{#6}}
\newcommand{\ottlongprodline}[2]{& & $#1$ & \multicolumn{4}{l}{$#2$}}
\newcommand{\ottfirstlongprodline}[2]{\ottlongprodline{#1}{#2}}
\newcommand{\ottbindspecprodline}[6]{\ottprodline{#1}{#2}{#3}{#4}{#5}{#6}}
\newcommand{\ottprodnewline}{\\}
\newcommand{\ottinterrule}{\\[5.0mm]}
\newcommand{\ottafterlastrule}{\\}
\newcommand{\ottmetavars}{
\ottmetavartabular{
 $ \mathit{id} $ &  \\
 $ \mathit{n} $ &  \\
 $ \ottmv{i} $ &  \\
 $ \ottmv{j} $ &  \\
 $ \ottmv{k} $ &  \\
 $ \ottmv{n} $ &  \\
 $ \ottmv{m} $ &  \\
 $ \ottmv{p} $ &  \\
 $ \ottmv{q} $ &  \\
}}

\newcommand{\ottidentifier}{
\ottrulehead{\mathit{id}}{::=}{}\ottprodnewline
\ottfirstprodline{|}{ {\sf \mathit{id} } }{}{}{}{}}

\newcommand{\ottRvar}{
\ottrulehead{\mathit{x}}{::=}{}\ottprodnewline
\ottfirstprodline{|}{ {\sf \mathit{id} } }{}{}{}{}}

\newcommand{\ottinstr}{
\ottrulehead{\ottnt{instr}}{::=}{\ottcom{}}\ottprodnewline
\ottfirstprodline{|}{ \textit{Binop} ~    ( \mathit{a}_{{\mathrm{1}}} ,~ \mathit{a}_{{\mathrm{2}}} )~ \mathit{env} }{}{}{}{\ottcom{binary op.}}\ottprodnewline
\ottprodline{|}{ \pirI{Branch}~   \mathit{L} }{}{}{}{\ottcom{jump}}\ottprodnewline
\ottprodline{|}{ \pirI{Branch}~  ( \mathit{a} ,~ \mathit{L}_{{\mathrm{1}}} ,~ \mathit{L}_{{\mathrm{2}}} ) }{}{}{}{\ottcom{branch}}\ottprodnewline
\ottprodline{|}{ \pirI{Call}~   \mathit{a}_{{\mathrm{0}}} ~( \mathit{a}^{*} )~ \mathit{env} }{}{}{}{\ottcom{apply closure}}\ottprodnewline
\ottprodline{|}{ \pirI{Deopt}~( \mathit{id} ,~ \mathit{a}^{*} ,~ \mathit{env} ) }{}{}{}{\ottcom{deoptimization}}\ottprodnewline
\ottprodline{|}{ \pirI{Force}~        ( \mathit{a} )~ \mathit{env} }{}{}{}{\ottcom{force promise}}\ottprodnewline
\ottprodline{|}{ \pirI{LdArg}~        ( \mathit{n} ) }{}{}{}{\ottcom{load argument}}\ottprodnewline
\ottprodline{|}{ \pirI{LdConst}~       \mathit{c} }{}{}{}{\ottcom{load constant}}\ottprodnewline
\ottprodline{|}{ \pirI{LdFun}~        ( \mathit{x} ,~ \mathit{env} ) }{}{}{}{\ottcom{load function}}\ottprodnewline
\ottprodline{|}{ \pirI{LdVar}~        ( \mathit{x} ,~ \mathit{env} ) }{}{}{}{\ottcom{load variable}}\ottprodnewline
\ottprodline{|}{ \pirI{MkArg}~    ( \mathit{id} ,~ \mathit{env} ) }{}{}{}{\ottcom{create promise}}\ottprodnewline
\ottprodline{|}{ \pirI{MkEnv}~  ( {(\mathit{x}=\mathit{a})}^{*} ~ : \mathit{env} ) }{}{}{}{\ottcom{create env.}}\ottprodnewline
\ottprodline{|}{ \pirI{MkClosure}~( \mathit{id} ,~ \mathit{env} ) }{}{}{}{\ottcom{create closure}}\ottprodnewline
\ottprodline{|}{ \pirI{Phi}~  ( {(\mathit{L} : \mathit{v})}^{*} ) }{}{}{}{\ottcom{$\phi$ function}}\ottprodnewline
\ottprodline{|}{ \pirI{Return}~       ( \mathit{a} ) }{}{}{}{\ottcom{return}}\ottprodnewline
\ottprodline{|}{ \pirI{StVar}~     ( \mathit{x} ,~ \mathit{a} ,~ \mathit{env} ) }{}{}{}{\ottcom{store variable}}}

\newcommand{\ottst}{
\ottrulehead{\ottnt{st}}{::=}{\ottcom{statements}}\ottprodnewline
\ottfirstprodline{|}{ ({\sf \%}n~|~{\sf e}n)  =  \ottnt{instr} }{}{}{}{\ottcom{non-void instruction}}\ottprodnewline
\ottprodline{|}{ \ottnt{instr} }{}{}{}{\ottcom{void instruction}}}

\newcommand{\ottargs}{
\ottrulehead{\mathit{a}^{*}}{::=}{}\ottprodnewline
\ottfirstprodline{|}{\mathit{a}_{{\mathrm{1}}}  \ottsym{,} \, .. \, \ottsym{,}  \mathit{a}_{\ottmv{n}}}{}{}{}{}}

\newcommand{\ottbindings}{
\ottrulehead{{(\mathit{x}=\mathit{a})}^{*}}{::=}{}\ottprodnewline
\ottfirstprodline{|}{\ottnt{binding_{{\mathrm{1}}}}  \ottsym{,} \, .. \, \ottsym{,}  \ottnt{binding_{\ottmv{n}}}}{}{}{}{}}

\newcommand{\ottbinding}{
\ottrulehead{\ottnt{binding}}{::=}{}\ottprodnewline
\ottfirstprodline{|}{\mathit{x}  \ottsym{=}  \mathit{a}}{}{}{}{}}

\newcommand{\ottscoping}{
\ottrulehead{: \mathit{env}}{::=}{}\ottprodnewline
\ottfirstprodline{|}{ :  \mathit{env} }{}{}{}{}}

\newcommand{\ottBinop}{
\ottrulehead{\textit{Binop}}{::=}{}\ottprodnewline
\ottfirstprodline{|}{ \pirI{Add} }{}{}{}{}\ottprodnewline
\ottprodline{|}{ ... }{}{}{}{}}

\newcommand{\ottBBId}{
\ottrulehead{\mathit{L}}{::=}{}\ottprodnewline
\ottfirstprodline{|}{ \mathsf{BB}_ \mathit{n} }{}{}{}{}}

\newcommand{\ottphiInputs}{
\ottrulehead{{(\mathit{L} : \mathit{v})}^{*}}{::=}{}\ottprodnewline
\ottfirstprodline{|}{\ottnt{phiInput_{{\mathrm{1}}}}  \ottsym{,} \, .. \, \ottsym{,}  \ottnt{phiInput_{\ottmv{n}}}}{}{}{}{}}

\newcommand{\ottphiInput}{
\ottrulehead{\ottnt{phiInput}}{::=}{}\ottprodnewline
\ottfirstprodline{|}{ \mathit{L}  :  \mathit{a} }{}{}{}{}}

\newcommand{\ottarg}{
\ottrulehead{\mathit{a}}{::=}{}\ottprodnewline
\ottfirstprodline{|}{({\sf \%}n~|~{\sf e}n)}{}{}{}{}\ottprodnewline
\ottprodline{|}{\ottnt{lit}}{}{}{}{}}

\newcommand{\ottargOrEnv}{
\ottrulehead{\mathit{a}, \mathit{env}}{::=}{\ottcom{argument}}\ottprodnewline
\ottfirstprodline{|}{({\sf \%}n~|~{\sf e}n)}{}{}{}{\ottcom{register}}\ottprodnewline
\ottprodline{|}{\ottnt{lit}}{}{}{}{\ottcom{literal}}}

\newcommand{\ottconst}{
\ottrulehead{\mathit{c}}{::=}{}\ottprodnewline
\ottfirstprodline{|}{ [  \mathit{n}  ]~ \ottnt{vecP} }{}{}{}{}}

\newcommand{\ottvecP}{
\ottrulehead{\ottnt{vecP}}{::=}{}\ottprodnewline
\ottfirstprodline{|}{\mathit{n} \, .. \, \mathit{n}}{}{}{}{}}

\newcommand{\ottans}{
\ottrulehead{({\sf \%}n~|~{\sf e}n)}{::=}{}\ottprodnewline
\ottfirstprodline{|}{\mathtt{\%}\mathit{n}}{}{}{}{}\ottprodnewline
\ottprodline{|}{\mathtt{e}\mathit{n}}{}{}{}{}}

\newcommand{\ottvarName}{
\ottrulehead{\mathtt{\%}\mathit{n}}{::=}{}\ottprodnewline
\ottfirstprodline{|}{ \mathtt{\% \mathit{n} } }{}{}{}{}\ottprodnewline
\ottprodline{|}{ \mathtt{\%}\mathit{i} }{}{}{}{}\ottprodnewline
\ottprodline{|}{ \mathtt{\%}\mathit{j} }{}{}{}{}\ottprodnewline
\ottprodline{|}{ \mathtt{\%}\mathit{k} }{}{}{}{}\ottprodnewline
\ottprodline{|}{ \mathtt{\%}\mathit{n} }{}{}{}{}\ottprodnewline
\ottprodline{|}{ \mathtt{\%}\mathit{m} }{}{}{}{}\ottprodnewline
\ottprodline{|}{ \mathtt{\%}\mathit{p} }{}{}{}{}\ottprodnewline
\ottprodline{|}{ \mathtt{\%}\mathit{q} }{}{}{}{}\ottprodnewline
\ottprodline{|}{ \mathtt{\%}\mathit{1}, \dots, \mathtt{\%}\mathit{n} }{}{}{}{}}

\newcommand{\ottenvVarName}{
\ottrulehead{\mathtt{e}\mathit{n}}{::=}{}\ottprodnewline
\ottfirstprodline{|}{ \mathtt{e{ \mathit{n} } } }{}{}{}{}\ottprodnewline
\ottprodline{|}{ \mathtt{e}\mathit{i} }{}{}{}{}\ottprodnewline
\ottprodline{|}{ \mathtt{e}\mathit{j} }{}{}{}{}\ottprodnewline
\ottprodline{|}{ \mathtt{e}\mathit{k} }{}{}{}{}\ottprodnewline
\ottprodline{|}{ \mathtt{e}\mathit{n} }{}{}{}{}\ottprodnewline
\ottprodline{|}{ \mathtt{e}\mathit{m} }{}{}{}{}\ottprodnewline
\ottprodline{|}{ \mathtt{e}\mathit{p} }{}{}{}{}\ottprodnewline
\ottprodline{|}{ \mathtt{e}\mathit{q} }{}{}{}{}}

\newcommand{\ottenv}{
\ottrulehead{\mathit{env}}{::=}{\ottcom{NOT THE ACTUAL ENV DEFINITON}}\ottprodnewline
\ottfirstprodline{|}{\mathtt{e}\mathit{n}}{}{}{}{}\ottprodnewline
\ottprodline{|}{\mathsf{G}}{}{}{}{}\ottprodnewline
\ottprodline{|}{\mathsf{O}}{}{}{}{}}

\newcommand{\ottlit}{
\ottrulehead{\ottnt{lit}}{::=}{\ottcom{literals}}\ottprodnewline
\ottfirstprodline{|}{\mathsf{G}}{}{}{}{\ottcom{global env.}}\ottprodnewline
\ottprodline{|}{\mathsf{O}}{}{}{}{\ottcom{placeholder env.}}\ottprodnewline
\ottprodline{|}{ \_ }{}{}{}{\ottcom{no value}}\ottprodnewline
\ottprodline{|}{ true }{}{}{}{\ottcom{true}}\ottprodnewline
\ottprodline{|}{ ... }{}{}{}{}}

\newcommand{\ottC}{
\ottrulehead{\ottnt{C}}{::=}{\ottcom{Code}}\ottprodnewline
\ottfirstprodline{|}{\mathit{B^*}}{}{}{}{}}

\newcommand{\ottP}{
\ottrulehead{\ottnt{P}}{::=}{\ottcom{Promise}}\ottprodnewline
\ottfirstprodline{|}{\mathit{id}  \ottsym{:}  \ottnt{C}}{}{}{}{}}

\newcommand{\ottF}{
\ottrulehead{\ottnt{F}}{::=}{}\ottprodnewline
\ottfirstprodline{|}{\ottnt{F}  \ottsym{:}  \ottnt{V_{{\mathrm{1}}}}  \ottsym{,} \, .. \, \ottsym{,}  \ottnt{V_{\ottmv{n}}}}{}{}{}{}}

\newcommand{\ottV}{
\ottrulehead{\ottnt{V}}{::=}{\ottcom{Function}}\ottprodnewline
\ottfirstprodline{|}{\ottnt{C} \, \mathit{P^*}}{}{}{}{\ottcom{body and promises}}}

\newcommand{\ottPs}{
\ottrulehead{\mathit{P^*}}{::=}{}\ottprodnewline
\ottfirstprodline{|}{\ottnt{P_{{\mathrm{1}}}} \, .. \, \ottnt{P_{\ottmv{n}}}}{}{}{}{}}

\newcommand{\ottB}{
\ottrulehead{\ottnt{B}}{::=}{\ottcom{Basic Block}}\ottprodnewline
\ottfirstprodline{|}{ ~~ \mathit{L} :  \mathit{ {st}^{*} } }{}{}{}{}}

\newcommand{\ottBBBody}{
\ottrulehead{\mathit{ {st}^{*} }}{::=}{}\ottprodnewline
\ottfirstprodline{|}{\ottnt{stP_{{\mathrm{1}}}} \, .. \, \ottnt{stP_{\ottmv{n}}}}{}{}{}{}}

\newcommand{\ottstP}{
\ottrulehead{\ottnt{stP}}{::=}{}\ottprodnewline
\ottfirstprodline{|}{ &  ({\sf \%}n~|~{\sf e}n)  & = &  \ottnt{instr}  \\ }{}{}{}{}\ottprodnewline
\ottprodline{|}{ &  ({\sf \%}n~|~{\sf e}n)  & = & ... \\ }{}{}{}{}\ottprodnewline
\ottprodline{|}{ &        &   &  \ottnt{instr}  \\ }{}{}{}{}\ottprodnewline
\ottprodline{|}{ &        &   & ... \\ }{}{}{}{}\ottprodnewline
\ottprodline{|}{ & \multicolumn{3}{l}{\texttt{\#  \ottnt{comment} } } \\ }{}{}{}{}}

\newcommand{\ottprintableProgram}{
\ottrulehead{\ottnt{printableProgram}}{::=}{}\ottprodnewline
\ottfirstprodline{|}{ {\ttfamily\small
                 \vskip -0.5em
                  \[
                      \begin{array}{rlll}
                          \multicolumn{4}{l}{
                              \ottnt{printableWithLabel} 
                          }
                      \end{array}
                  \]
                 \vskip -0.01em
              } }{}{}{}{}\ottprodnewline
\ottprodline{|}{ {\ttfamily\small
                 \vskip -0.5em
                  \[
                      \begin{array}{rlll}
                          \ottnt{printable} 
                      \end{array}
                  \]
                 \vskip -0.01em
              } }{}{}{}{}}

\newcommand{\ottprintableWithLabel}{
\ottrulehead{\ottnt{printableWithLabel}}{::=}{}\ottprodnewline
\ottfirstprodline{|}{\ottnt{namedPrintableFunction_{{\mathrm{1}}}} \, .. \, \ottnt{namedPrintableFunction_{\ottmv{n}}}}{}{}{}{}}

\newcommand{\ottBs}{
\ottrulehead{\mathit{B^*}}{::=}{}\ottprodnewline
\ottfirstprodline{|}{\ottnt{B_{{\mathrm{1}}}} \, .. \, \ottnt{B_{\ottmv{n}}}}{}{}{}{}}

\newcommand{\ottprintable}{
\ottrulehead{\ottnt{printable}}{::=}{}\ottprodnewline
\ottfirstprodline{|}{\mathit{ {st}^{*} }}{}{}{}{}\ottprodnewline
\ottprodline{|}{\mathit{B^*}}{}{}{}{}}

\newcommand{\ottnamedPrintableFunction}{
\ottrulehead{\ottnt{namedPrintableFunction}}{::=}{}\ottprodnewline
\ottfirstprodline{|}{ } \\  \mathit{ {st}^{*} }  \multicolumn{4}{l}{ }{}{}{}{}\ottprodnewline
\ottprodline{|}{ \mathsf{ \mathit{id} } } \\  \mathit{ {st}^{*} }  \multicolumn{4}{l}{ }{}{}{}{}\ottprodnewline
\ottprodline{|}{ \mathsf{ \mathit{id} } } \\  \mathit{B^*} \multicolumn{4}{l}{ }{}{}{}{}}

\newcommand{\ottcomment}{
\ottrulehead{\ottnt{comment}}{::=}{}\ottprodnewline
\ottfirstprodline{|}{\mathit{id} \, \ottnt{commentPart_{{\mathrm{1}}}} \, .. \, \ottnt{commentPart_{\ottmv{n}}}}{}{}{}{}}

\newcommand{\ottcommentPart}{
\ottrulehead{\ottnt{commentPart}}{::=}{}\ottprodnewline
\ottfirstprodline{|}{ \kern-0.68em  \mathit{id} }{}{}{}{}}

\newcommand{\ottgrammar}{\ottgrammartabular{
\ottidentifier\ottinterrule
\ottRvar\ottinterrule
\ottinstr\ottinterrule
\ottst\ottinterrule
\ottargs\ottinterrule
\ottbindings\ottinterrule
\ottbinding\ottinterrule
\ottscoping\ottinterrule
\ottBinop\ottinterrule
\ottBBId\ottinterrule
\ottphiInputs\ottinterrule
\ottphiInput\ottinterrule
\ottarg\ottinterrule
\ottargOrEnv\ottinterrule
\ottconst\ottinterrule
\ottvecP\ottinterrule
\ottans\ottinterrule
\ottvarName\ottinterrule
\ottenvVarName\ottinterrule
\ottenv\ottinterrule
\ottlit\ottinterrule
\ottC\ottinterrule
\ottP\ottinterrule
\ottF\ottinterrule
\ottV\ottinterrule
\ottPs\ottinterrule
\ottB\ottinterrule
\ottBBBody\ottinterrule
\ottstP\ottinterrule
\ottprintableProgram\ottinterrule
\ottprintableWithLabel\ottinterrule
\ottBs\ottinterrule
\ottprintable\ottinterrule
\ottnamedPrintableFunction\ottinterrule
\ottcomment\ottinterrule
\ottcommentPart\ottafterlastrule
}}

\newcommand{\ottdefnss}{
}

\newcommand{\ottall}{\ottmetavars\\[0pt]
\ottgrammar\\[5.0mm]
\ottdefnss}